\documentclass[12pt,titlepage]{article}
\usepackage{epsfig}
\usepackage{graphicx}
\usepackage{amsmath}
\usepackage{color}

\def\beq{\begin{equation}}
\def\eeq{\end{equation}}
\def\bea{\begin{eqnarray}}
\def\eea{\end{eqnarray}}

\begin{document} 

\title{Pressure and compressibility of conformal field theories from the AdS/CFT correspondence}
\author{{Brian P. Dolan}\\
{\small Dept. of Mathematical Physics, 
Maynooth University, Maynooth, Ireland}\\
{\small Dublin Institute for Advanced Studies, 10 Burlington Rd, Dublin, Ireland}\\
{\small Email:} {\tt bdolan@thphys.nuim.ie}
}

\maketitle

\abstract{The equation of state associated with ${\cal N}=4$ supersymmetric Yang-Mills in 4 dimensions, for $SU(N)$ in the large $N$ limit, is investigated using the AdS/CFT correspondence. An asymptotically AdS black-hole on the gravity side provides a thermal background for the Yang-Mills theory on the boundary in which the cosmological constant is equivalent to a volume. The thermodynamic variable conjugate to the cosmological constant is a pressure and the $P-V$ diagram is studied. It is known that there is a critical point where the heat capacity diverges 
and this is reflected in the isothermal compressibility.  Critical exponents are derived and found to be mean field in the large $N$ limit. The same analysis applied to 3 and 6 dimensional conformal field theories again yields mean field exponents associated with the compressibility at the critical point.\footnote{Feature invited paper for special edition \lq\lq Black hole thermodynamics II'' 
in {\it Entropy}.}\\
\\
\rightline{DIAS-STP-16-05}
}

\newpage
%\leftline{PACS nos: 04.50.Gh; 04.70.Dy; 05.70.-a \hfill DIAS-STP-2015-01}

\section{Introduction}

Ever since Bekenstein's discovery of the relation between entropy and 
the area of a black hole's event horizon \cite{Bekenstein} and Hawking's subsequent observation that black holes have an intrinsic temperature
associated with them \cite{Hawking}, the subject of black hole thermodynamics has been a fascinating area of research. Indeed black hole thermodynamics now has
potentially great usefulness in providing insights into the thermodynamics of conformal theories via the AdS/CFT correspondence \cite{ABJM}.

An interesting aspect of black hole thermodynamics is the role of the cosmological constant, $\Lambda$, as a thermodynamic variable: an idea originally considered in \cite{Henneaux} and re-visited in \cite{KRT}. Since $\Lambda$ has the natural physical interpretation of pressure in the bulk, the conjugate variable is a volume and this has led to some intriguing recent results for the pressure and volume of black holes, with phase transitions and critical points showing
very similar properties to liquid-gas phase transitions \cite{Galaxies,LambdaReview}. 
For Einstein gravity in the bulk, the critical exponents are mean field for all black hole solutions studied to date, leading to an analogy with the liquid-gas transition of a van der Waals fluid.

In the context of AdS/CFT it was suggested in \cite{KRT,CJ} that varying the cosmological constant in the bulk should be associated with varying the number of colors in the boundary CFT and this proposal was investigated in more detail in \cite{BCB}, where the thermodynamically conjugate variable to $\Lambda$ was interpreted as a kind of chemical potential for color.

An alternative approach is followed here based on a different interpretation of the thermodynamic significance of $\Lambda$ which provides a length scale in the CFT.
When global AdS co-ordinates are used the CFT on the boundary has a finite volume, determined by $\Lambda$, hence varying $\Lambda$ in the bulk corresponds to varying the volume of the CFT on the boundary. If $\Lambda$ gives the volume of the CFT then the thermodynamically conjugate variable is the pressure and one is led to the construction of $P-V$ diagrams for the CFT, in which the pressure and volume are interchanged relative to their roles in the bulk.
This can be achieved while keeping the number of colors fixed provided the higher dimensional Newton's constant is adjusted as $\Lambda$ is varied, as suggested in \cite{KR}. In $10$-dimensional string theory, for example, Newton's constant is related to the string coupling constant and tension by $G_{10}=8 \pi^6 g_s^2(\alpha')^4$ so varying $G_{10}$ can be thought of as varying $g_s$ with the tension fixed.  

In this paper the $P-V$ diagram associated with
$4$-dimensional ${\cal N}=4$ SUSY $SU(N)$ Yang-Mills theory at large 
$N$ is investigated using the AdS/CFT correspondence.
When the bulk contains a black hole 
there is a first order phase transition \cite{HP}  which is 
associated with the deconfining transition for the quark-gluon plasma 
on the boundary \cite{Witten}.  When the black hole carries a $U(1)$ charge, 
corresponding to $R$-charge in the Yang-Mills theory, this phase transition
is one end of a line of first order phase transitions while the other end terminates at a second order transition where the heat capacity at constant volume
for the Yang-Mills theory diverges \cite{CEJM1}. 

The critical exponents in the bulk gravitational theory,
with charge as the order parameter and temperature as the control
parameter, are known to be mean field, \cite{NTW}, and so $C_V$ is finite at the critical point for the black hole in the bulk.
At first sight this is at odds with the statement in \cite{CEJM1} that there is a
critical point in the CFT at which $C_V$ diverges. 
We shall show that there is no contradiction here and, when interpreted correctly, the boundary CFT has mean field exponents. 
In terms of pressure and volume there is a number of aspects of the phase transition that make it different from more usual cases.  Firstly there is a single phase at low temperatures and the two phase regime exists for temperatures above the critical point \cite{CEJM1};
secondly, above the critical point, along the line of first order phase transitions, it is the pressure that jumps across the phase transition, not the volume, so it is more appropriate to use the pressure as an order parameter rather than the more usual volume; thirdly the conformal symmetry dictates that volume and temperature are not fully independent and it is better to use charge as the control parameter rather than the temperature.  This last point of view is more in keeping with the notion of the phase transition being a quantum phase transition rather than a thermal phase transition. With this interpretation the critical exponents for pressure and volume of the Yang-Mills theory are calculated
in the large $N$ limit and shown to be mean field.

The general structure is the same for the $3$-dimensional and $6$-dimensional CFT's considered in \cite{Maldacena}, in particular the critical exponents at large $N$ are mean field in all three conformal field theories.

In section \ref{sec:bulk} the black hole thermodynamics of the relevant bulk solution is summarized and is related to that of the boundary CFT in section \ref{CFT}. Section \ref{sec:YM} analyzes the case
of ${\cal N}=4$ SUSY Yang-Mills in detail, the $P-V$ diagram is constructed and
critical exponents for the deconfining phase transition in the large $N$ limit are calculated. Finally section \ref{summary} summarizes the results. Some technical details are in two appendices.
 
\section{The bulk}\label{sec:bulk}

In $D$ space-time dimensions the Lagrangian for Einstein gravity, with a cosmological constant $\Lambda$, coupled to a $U(1)$ gauge field, is
\beq \label{L_D}
{\cal L}_E = \frac{1}{16 \pi G_D} ( R - 2\Lambda - F^2).
\eeq  
The normalization of $F$ is the same as in \cite{CEJM1}, it has dimensions of inverse length. 

A solution of the equations of motion arising from (\ref{L_D}), corresponding a spherical charged asymptotically AdS black hole in $D$ space-time dimensions
is easily written down.
In global co-coordinates the gauge potential is
\beq \label{A-ansatz}
A = \frac{\widetilde Q}{(D-3)\Omega_{D-2}  }  \left(\frac{1}{r^{D-3} } - c_0  \right) d t 
\eeq
where $\Omega_{D-2}=\frac{2\pi^{\frac{D-1}{2}}}{\Gamma\left(\frac{D-1}{2}\right)}$ is the volume of a unit $(D-2)$-sphere.
The constant $c_0$ accommodates some freedom in the choice of gauge.

The normalization of the $U(1)$ charge $\widetilde Q$ here is determined by requiring that the gauge field $F=dA$ satisfies Gauss law
\beq
\widetilde Q = \int_{S^{D-2}}*F
\eeq
where $S^{D-2}$ is a sphere containing the charge.
 
The line element is
\beq \label{ds_squared}
d^2 s = -f(r) d t^2 + \frac{dr^2}{f(r)} + r^2 d^2\Omega_{D-2}, 
\eeq
where
\beq \label{fr}
f(r) = 1 - \frac{\mu}{r^{D-3}} + \frac{q^2}{r^{2(D-3)}} +\frac{r^2}{L^2},
\eeq
$\Lambda = -\frac{(D-1)(D-2)}{2 L^2}$ and $d^2\Omega_{D-2}$ is the line element on a unit $(D-2)$ sphere. 
There is an event horizon and the largest root of $f(r)=0$ will be denoted by $r_h$.  It is natural to chose a gauge in which the potential 
above vanishes at the outer horizon, $c_0=\frac{1}{r_h^{D-3}}$.

The parameters  $q$ and $\mu$ are then
related to $\widetilde Q$ and the mass $M$ of the black hole by
\beq \widetilde Q^2 = \frac{(D-2)(D-3)\Omega^2_{D-2} q^2}{2}
\eeq
and
\beq \label{M-mu}
M=\frac{(D-2)\Omega_{D-2} \mu}{16\pi G_D}\eeq
(we use units with $c=1$, but keep the $D$-dimensional Newton's constant explicit).

From (\ref{fr}), with $f(r_h)=0$, and (\ref{M-mu})
\beq
M=\frac{(D-2)\Omega_{D-2}}{16\pi G_D} \left(\frac{r_h^{D-1}}{L^2} + r_h^{D-3} +\frac{q^2}{r_h^{D-3}} \right).
\eeq

The area of the event horizon is $\Omega_{D-2} r_h^{D-2}$ and the Bekenstein-Hawking entropy is
\beq
S=\frac{\Omega_{D-2}}{4} \frac{r_h^{D-2}}{\hbar G_D}.
\eeq
From the point of view of black hole thermodynamics, the mass is interpreted as the internal energy of the system, $M=U(S,\widetilde Q)$, and the Bekenstein-Hawking
temperature is
\beq
T=\left.\frac{\partial U}{\partial S}\right|_{L,\widetilde Q}=\frac{\hbar}{4\pi r_h^{D-3}}\left\{(D-1)\frac{r_h^{D-2}}{L^2} + (D-3)r_h^{D-4} -(D-3)\frac{q^2}{r_h^{D-2}} \right\}.
\eeq

It will prove useful to define dimensionless variables, $x:=\frac{r_h}{L}$ and $y:=\frac{q}{L^{D-3}}$ which can be used in lieu of $S$ and the charge. In terms of these variables
\beq \label{Mx}
M(x,y)=\frac{(D-2)\Omega_{D-2} L^{D-3}}{16 \pi G_D}\left(x^{D-1} + x^{D-3} + \frac{y^2}{x^{D-3}} \right)
\eeq
and
\beq \label{Tx}
T(x,y)=\frac{\hbar}{4 \pi L} \left\{ (D-1)x + \frac{(D-3)}{x} -
(D-3)\frac{y^2}{x^{2D-5}}\right\}.
\eeq
 
In the context of superstring theory, the $D$-dimensional Newton's constant $G_D$ is descended from Newton's constant in the full ${\cal D}$-dimensional theory (${\cal D}=10$ or $11$). Compactifying the higher dimensional theory on a 
$({\cal D}-D)$-dimensional compact space ${\cal K}_{{\cal D}-D}$, with size $L$ , allows for a direct product ${\cal D}$-dimensional space-time ${\cal M}_D \times {\cal K}_{{\cal D}-D}$, where ${\cal M}_D$ is asymptotically $AdS_D$ with line-element (\ref{ds_squared}), and
\beq
\frac{1}{16\pi G_D} = 
\frac{Vol({{\cal K}_{{\cal D}-D}})}{16 \pi G_{\cal{D}}}=
C\frac{L^{{\cal D}-D}}{16\pi G_{\cal D}},
\eeq
where $C$ is a dimensionless number determined by the metric on 
${\cal K}_{{\cal D}-D}$.
 The AdS length scale $L$ is not intrinsic to the ${\cal D}$-dimensional action, but is merely a parameter in a classical solution of the ${\cal D}$-dimensional theory and it is perfectly reasonable to consider varying $L$ in the solution.
 
\section{Thermodynamics of the boundary field theory}\label{CFT}

In the AdS/CFT correspondence the boundary CFT theory has $d=D-2$ space 
dimensions.  With the global co-oordinates used in the previous section, the
$(d+1)$-dimensional space-time metric at fixed $r$ is 
\beq
d^2 s = -f(r) d t^2 + r^2 d^2\Omega_d 
\quad \underset{r\rightarrow\infty}{\longrightarrow} \quad 
\frac{r^2}{L^2}( -d t^2 + L^2 d^2\Omega_d), 
\eeq
which is conformal to a $(d+1)$-dimensional space-time with constant time
spatial volume
\beq
V = \Omega_d L^d, 
\eeq
and this can be interpreted as the spatial volume of the boundary conformal field theory.

The dimensionless ratio of the AdS length scale $L$ to the ${\cal D}$-dimensional Planck length in the bulk determines the number of degrees of freedom in the boundary field theory \cite{Maldacena}.

The most studied case is  ${\cal D}=10$ compactified on ${\cal K}_5=S^5$, with $D=5$.
the boundary CFT is then ${\cal N}=4$ SUSY $SU(N)$ Yang-Mills theory, when the classical solution in the bulk (\ref{ds_squared}) is relevant for large $N$.
This has $16 N^2$ degrees of freedom ($8N^2$ bosonic and $8N^2$ fermionic) and
the large $N$ limit is the weak gravity limit: 
\beq \label{GLN-equations}
\frac{L^8}{\hbar G_{10}} = \frac{2 N^2}{\pi^4}, 
\qquad  Vol(S^5)=\pi^3 L^5, \qquad  \frac{1}{\hbar G_5} = \frac{2 N^2}{\pi L^3}.
\eeq
$\hbar G_{10}$ is of course related to the $10$-dimensional Planck 
length, $\hbar G_{10}=l_{Pl}^8$ and, as usual, large $N$ is related to the classical limit of the bulk theory, in which $L>>l_{Pl}$.

The CFT is on a 3-sphere with volume
\beq \label{Volume}
V=2\pi^2 L^3
\eeq 
and the inverse of the 5-dimensional Planck length (\ref{GLN-equations}) is related to the number of degrees of freedom per unit volume of the CFT 
\beq
\frac{1}{4 \pi \hbar G_5} = \frac{N^2}{V},
\eeq
so varying the volume of the CFT, keeping $N^2$ fixed, 
is completely equivalent to varying the 5-dimensional Planck length.

The entropy is proportional to $N^2$ (with $x:=\frac{r_h}{L}$ as before)
\beq
S = \pi N^2 x^3.
\eeq
as is the dimensionless charge
\beq \label{Qy}
Q=\frac{\widetilde Q L}{4\pi \hbar G_5} = \sqrt{3} N^2 y,
\eeq
where $y:=\frac{q}{L^2}$.

On dimensional grounds the internal energy $U(S,V,Q)=M$ is of the form 
\beq U(S,V,Q)= \frac{\hbar}{V^{\frac{1}{3}}}u(S,Q), 
\eeq
with $u(S,Q)$ a dimensionless function of dimensionless variables.
The thermodynamics of the CFT is determined by the functional dependence of the mass on $S$, $Q$ and $V$ (or equivalently, $x$, $y$ and $L$), explicitly 
\beq \label{Mx4}
M=\frac{3}{4}\frac{N^2 \hbar}{L}\left(x^4 + x^2 + \frac{y^2}{x^2}\right).
\eeq

Two other interesting cases are the 3-dimensional CFT obtained from
M-theory in ${\cal D}=11$ compactified on $AdS_4\times S^7$ 
(more generally $S^7/k$ for integral $k$, \cite{ABJM,CEJM1}) and the 6-dimensional CFT obtained from M-theory in
${\cal D}=11$ compactified on $AdS_7\times S^4$.  For these two cases the
$N^2$ behavior of the \lq\lq extensive'' quantities $M$, $S$ and $Q$ is 
replaced with $N^{3/2}$ and $N^3$ respectively and the powers of $x$ change as
in equations (\ref{Mx}) and (\ref{Tx}) but otherwise the analysis of the thermodynamics is similar to the case ${\cal D}=10$ and $D=5$ studied in \S\ref{sec:YM} below.

\section{The thermodynamics of  ${\cal N}=4$ SUSY Yang-Mills}\label{sec:YM}

To explore the thermodynamics of the CFT fully we wish to fix $N$ and allow
$S$, $Q$ and $V$ to vary. This means we must vary $L$, but at the same time
vary the ${\cal D}$-dimensional Newton constant in such a way that $N^2$ in
is fixed in the first equation of (\ref{GLN-equations}), \cite{KR}.

For concreteness we shall focus on ${\cal D}=D=5$, the general structure is the same the other two cases, in particular the critical exponents are the same.  
Thermodynamically we interpret $M=U(S,Q,V)$ 
in (\ref{Mx4}) as the internal energy of the large $N$ Yang-Mills theory at
strong coupling,
\beq
U=\frac{3}{4}\frac{N^2 \hbar}{L}\left(x^4 + x^2 + \frac{y^2}{x^2}\right).
\eeq
Note that, while the mass in the bulk is classical, the internal energy of the CFT is quantum mechanical and vanishes as $\hbar\rightarrow 0$ with $x$ and $y$ fixed.

The temperature is then
\beq \label{Temperature}
T= \left.\frac{\partial U}{\partial S}\right|_{Q,V}
=\frac{\hbar}{2\pi L }\left(2 x + \frac 1 x - \frac{y^2}{x^5}\right),
\eeq
in the deconfined phase this is interpreted as the temperature of the quark-gluon plasma \cite{Witten,CEJM1}.

The pressure is \cite{KR}
\beq \label{Pressure}
P= -\left.\frac{\partial U}{\partial V}\right|_{S,Q}
=\frac{N^2 \hbar}{8\pi^2 L^4}\left(x^4 + x^2 + \frac{y^2}{x^2}\right)
=\frac{\epsilon}{3}
\eeq
where $\epsilon=\frac{U}{V}$ is the energy density.
Hence 
\beq
U= 3 P V
\eeq
and the speed of sound is given by
\beq
\left.\frac{\partial \epsilon}{\partial P}\right|_{S,Q}=\frac{1}{3},
\eeq
simple consequences of the fact that dimensionally,
$U\propto \frac {1} {V^{1/3}}$.

Lastly the chemical potential is
\beq \label{ChemicalPotential}
\Phi= \left.\frac{\partial U}{\partial Q}\right|_{S,V}
=\frac{\sqrt{3} \hbar}{2 L}\frac{y}{x^2}.
\eeq
 
These expressions simplify at high $T$ (large $x$)
when the event horizon curvature and the charge are negligible.
This is deep into the de-confined phase of the quark-gluon plasma and in this 
limit
\beq S=\frac{N^2 \pi^2}{2\hbar^3} V T^3, \qquad
P=\frac{N^2 \pi^2 T^4}{8\hbar^3} \qquad \hbox{and} \qquad \Phi= \frac{Q \hbar^3}{V N^2 T^2}.
\eeq
Since $T\propto \hbar$ we see that the pressure and the chemical potential here are of quantum mechanical origin.

 At fixed charge and temperature there is a phase transition \cite{CEJM1}, an extension of the $\widetilde Q=0$ 
Hawking-Page phase transition in the bulk \cite{HP}, which is interpreted as the
deconfining phase transition in the boundary Yang-Mills theory \cite{Witten}.
The heat capacity 
\beq C_{V,Q}=T\left.\frac{\partial S}{\partial T}\right|_{V,Q}
\eeq
diverges when 
$\left.\frac{\partial T}{\partial S}\right|_{V,Q}=\left.\frac{\partial^2 U}{\partial S^2}\right|_{V,Q}$ vanishes and there
is a critical point when 
$\left.\frac{\partial^2 U}{\partial S^2}\right|_{V,Q}=\left.\frac{\partial^3 U}{\partial S^3}\right|_{V,Q}=0$. This happens at 
\beq \label{xcrit_ycrit}
x^2=x^2_*=\frac 1 {3},\qquad  y^2=y^2_*= \frac 1 {135},
\eeq
a critical point that was first found in \cite{CEJM1}.
The critical temperature is given by
\beq L T_* = \frac{4\sqrt{3}}{5 \pi}.\eeq
Conformal invariance dictates that only the combination $V^{\frac{1}{3}}T$ is determined at critical point, $V$ and $T$ are not fixed separately.\footnote{Finite temperature field theory is associated with periodicity in Euclidean time with period $1/T$ and 
Euclidean time parameterizes a circle $S^1$ of radius $\frac 1 {2\pi T}$. The spatial geometry is $S^3$ with radius $L$ and, because the theory is conformal, the physics can only depend on the ratio of the radii of these two spheres, namely $2\pi T L$.  The physics depends on $V^{\frac{1}{3}} T$, not on $V$ and $T$ separately, \cite{Witten}.}

The specific heat critical exponent $\alpha$ can easily be determined by fixing $y=y_*$ and expanding $1/C_{V,Q}$ around $x_*$.  With $x=(1+\varepsilon)/\sqrt{3}$
\beq
\frac{1}{C_{V,Q}(x,y_*)} \propto \varepsilon^2
\eeq
$\alpha$ can be extracted by similarly expanding the reduced temperature
at constant volume,
\beq
t=\frac{T-T_*}{T_*},\eeq
which gives
\beq
t \propto \varepsilon^3.
\eeq
The critical exponent follows from $C_{V,Q}\sim |t|^{-\alpha}$ with
\beq
\alpha = \frac{2}{3}.
\eeq

This singularity in the heat capacity was not found in the bulk thermodynamics of higher dimensional asymptotically AdS charged black holes studied in \cite{NTW,KM}, these authors found a critical point in the bulk but the exponents were mean field and $C_{V,Q}$ is finite there.
To understand this apparent contradiction we write
the full equation of state  in terms of $Q$ and $\Phi$.  Using (\ref{Qy})
and (\ref{ChemicalPotential}) 
to eliminate $x$ in favor of $Q$ and $\Phi$ in (\ref{Temperature}) gives
the equation of state
\beq T=\frac{N}{3\pi L}\left(\frac{3\frac{Q}{N^2} + 3L \Phi - 4 (L \Phi)^3}{\sqrt{\vline height 9pt width 0pt 2 L \Phi Q}}\right).\eeq
Expanding about the critical point (\ref{xcrit_ycrit}) 
with $y=y_*(1+\eta)$, $T=T_*(1+t)$ and $\Phi=\Phi_*(1+\varphi)$, where $L\Phi_*:=\frac{1}{2\sqrt{5}}$, leads to
\beq \label{EoS}
t=-\frac{\eta}{24}\left( 2 + 5 \varphi \right) -\frac{5}{48}\varphi^3 + \hbox{o}(\varphi^4,\eta\varphi^2,\eta^2).
\eeq
Aside from a change in the sign  of $t$ this has the same analytic structure as the equation of state for a van der Waals gas \cite{NTW} which, in reduced variables $p=\frac{P-P_*}{P_*}$ and $v=\frac{V-V_*}{V_*}$, reads
\beq \label{vdW}
t =\frac{p}{8}  \left( 2 + 3  v\right) +\frac 3 8 v^3 +\hbox{o}(v^4,p v^2).
\eeq
The van der Waals equation of state has mean field exponents, so one expects 
$\alpha=0$.
The previous calculation gave $\alpha = \frac 2 3$ because $Q$ was held fixed
 (fixed $\eta$) but the order parameter for the liquid-gas phase transition in the van der Waals gas is $v$, which would be analogous to $\varphi$ in (\ref{EoS}).
Indeed $C_{V,\Phi}$ is perfectly finite, \cite{CEJM1},
so $\alpha$ is indeed zero for fixed $\Phi$, which is the analogue of the van der Waals case.

If $Q$ is the order parameter the usual exponents are
defined by 
\beq
\eta \sim |t|^\beta, \qquad \varphi \sim |\eta|^\delta
\eeq 
and the response function, 
\beq\chi_T = \left.\frac{\partial Q}{\partial \Phi}\right|_T,
\eeq
behaves as
\beq \label{chiT}
\chi_T \propto - |t|^{-\gamma}
\eeq
when $Q=Q_* = \frac{N^2}{3\sqrt{5}}$.

The equation of state (\ref{EoS}) results in
\beq
\beta =1, \qquad \gamma=-\frac 2 3\qquad\hbox{and}
\qquad \delta = \frac 1 3.
\eeq
While these exponents, together with $\alpha=\frac 2 3$, 
do satisfy the Rushbrooke and Widom scaling relations
\beq \alpha + 2\beta + \gamma =2, \qquad  \gamma=\beta(\delta -1)\eeq
there is a difficulty in that $\gamma$ is negative
so the response function vanishes at critically rather than diverging.
As indicated in (\ref{chiT}) the response function 
$\chi_T$ can be negative near the critical point,    
\beq
\left.\frac{\partial \eta}{\partial \varphi}\right|_t
\approx -\frac{5(2\eta + 3\varphi^2)}{4} < 0 \quad \hbox{for}\quad \eta =0,
\eeq
and this is the instability found in \cite{CEJM2}.\footnote{The same instability is also present for rotating Myers-Perry black holes with fixed angular momenta, 
which have a second order phase transition at
a critical value of the angular momenta \cite{CCK, Stability}.}
It was also shown in \cite{CEJM2} that $\Phi$ jumps across the phase transition in the two phase regime, suggesting that $\Phi$ is the order parameter for this transition. 

If $Q$ is just fixed from the start and not varied then there is no problem,
$C_{V,Q}$ diverges with exponent $\frac 2 3$ and $\beta$, $\gamma$ and $\delta$ are not relevant.  But if we wish to probe the system by adjusting $Q$ then 
$\Phi$ should be viewed as the order parameter, not $Q$. 
In that case $C_{V,Q}$ diverges as $|t|^{-1}$ rather than
$|t|^{-\frac 2 3}$ because it should be calculated for $\varphi=0$ and not $\eta=0$. 

The other exponents, $\varphi \sim |t|^\beta$ and $\varphi \sim |\eta|^\delta$, 
are easily obtained from (\ref{EoS}) in the usual way and are found to be mean field 
\beq
\alpha=0,\qquad \beta =\frac 1 2, \qquad \gamma=1 \qquad\hbox{and}
\qquad \delta = 3.
\eeq
In particular the heat capacity $C_{V,\Phi}$ is finite (and negative) at the critical point \cite{CEJM1}.  Mean field behavior with $\Phi$ as the order parameter for the black hole in the bulk was first found in \cite{NTW}.

Alternatively the equation of state can be written 
in terms of the pressure rather than the chemical potential to study the compressibility.
The adiabatic compressibility of the plasma follows easily from (\ref{Pressure}) and is
\beq \label{kappaS}
\kappa_{S,Q}=-\frac{1}{V}\left.\frac{\partial V}{\partial P}\right|_{S,Q}
=\frac {3}{4 P},
\eeq
but the isothermal properties are more interesting.  

It is shown in appendix \ref{AA} that the isothermal compressibility, 
$\kappa_{T,Q}$, is given by
\beq \label{kappaT}
\kappa_{T,Q} = \frac{9 V } {12 P -T C_{V,Q}}.
\eeq
Hence $\kappa_{T,Q}$ is vanishes at the critical point and is negative
along $Q=Q_*$  close to the critical point when $C_{V,Q_*}>0$ diverges.  Thus the $P$-$V$ response function, $\kappa_{T,Q}$, behaves the same way as the $Q$-$\Phi$ response function
$\chi_T$. This suggests that the
order parameter here is $P$ rather than $V$.

If this is a valid point of view we should focus on
the heat capacity at constant pressure, $C_{P,Q}$, rather than $C_{V,Q}$.
Using the standard relation
\beq
\kappa_S C_P = \kappa_T C_V
\eeq
equations (\ref{kappaS}) and (\ref{kappaT}) give
\beq
C_{P,Q}= \frac{12 PV C_{V,Q}}{12 P - T C_{V,Q}}\eeq
which is finite and negative when $C_{V,Q}$ diverges, just as $C_{V,\Phi}$ is.
In particular at the critical point
\beq C_{P,Q}=  -12 \left(\frac{PV}{T}\right)_* = -\frac{7 \pi N^2}{4\sqrt{3}}.
\eeq
Thus, with $P$ as the order parameter, $\alpha=0$ is mean field.
The negative value of $C_P$ at the critical point is a reflection of the instability found above in $\chi_T$.

There is a subtlety in trying to extract $\beta$ and $\delta$ as there is no
independent definition of a critical volume, $V$ and $T$ are linked due to conformal invariance.  In particular it would be wrong to assume the usual scaling
relations to derive $\beta$ and $\delta$ from $\alpha$ and $\gamma$ above
as conformal invariance imposes an extra constraint.

A reduced volume can be defined as
\beq \label{VT3}
v:=\frac{V T^3-(VT^3)_*}{(VT^3)_*}
\eeq
but this is not a new variable since conformal invariance relates it directly to $t$,
\beq \label{v3t} 1+v = (1+t)^3.
\eeq
As a consequence the usual definitions of the exponents $\beta$ and $\delta$
do not apply.  

\begin{figure}[ht]
\centerline{{\includegraphics[scale=1.3]{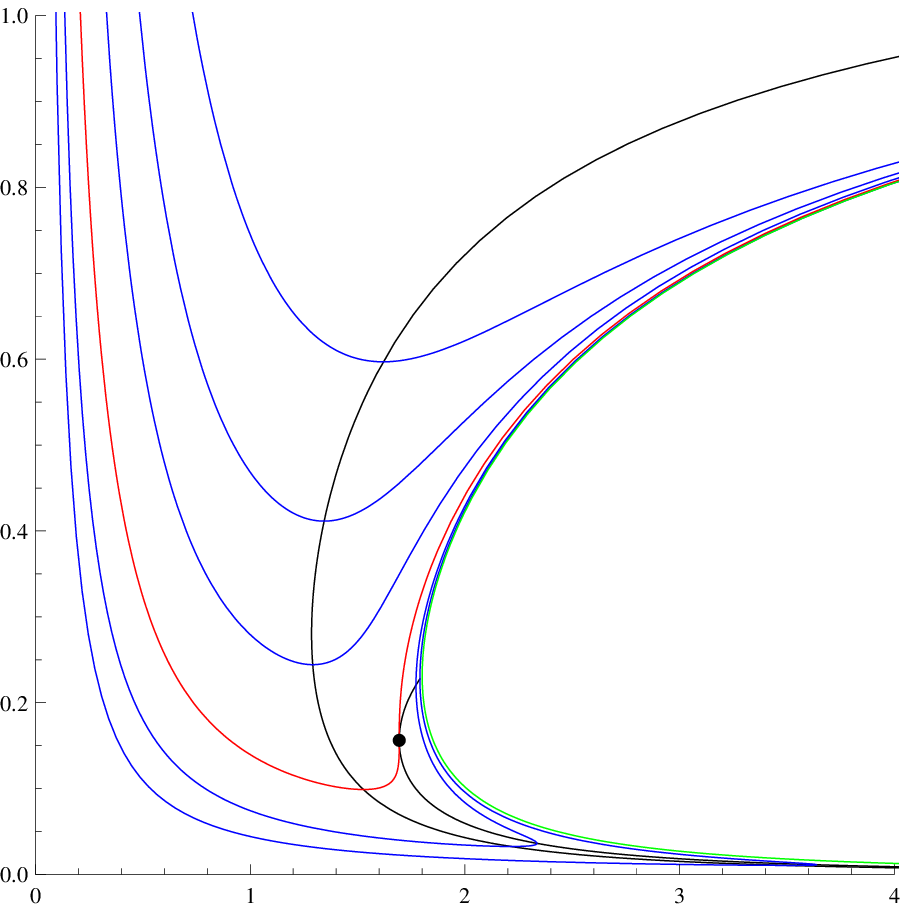}}}
\caption{Curves of constant $Q$ in the $P-V$ plane ($P/ N^2 T^4$ plotted against $VT^3$).
The isothermal compressibility is negative in the region between the two black curves.  The rightmost (green) curve is $Q=0$ and the red curve is $Q=Q_*$.  
The critical point is indicated by the black dot, it lies on the boundary of the small right-hand region in which $\kappa_{T,Q}$ is positive, between the right-hand black curve and the green curve. $C_P>0$ only to the left of the left-hand black curve. 
}\label{PV-figure}
\end{figure}

Since $v$ and $t$ are not independent it is better, when discussing
pressure and volume, to use $\eta$ 
to probe the physics near the critical point rather than $t$.  
Lines of equal charge in the $P$-$V$ plane are plotted in figure
\ref{PV-figure} where $P$ and $V$ are rendered dimensionless by multiplying by appropriate powers of $T$. The isothermal compressibility is positive when the slope of the plotted curves is negative and there are two regions in the $P$-$V$ plane where this is the case, the critical point lies on the boundary of the rightmost of these regions which corresponds to $Q<Q_*$ and $v>0$ 
(hence $t>0$, {\it i.e.} the high temperature regime). The pressure of the system jumps
across the phase transition.

To calculate critical exponents first define a reduced pressure
\beq p = \frac{P-P_*}{P_*}\eeq 
at fixed volume.\footnote{One can also use the temperature to render $P$ dimensionless and define the reduced pressure as $p=\frac{P/T^4 -(P/T^4)_*}{(P/T^4)_*}$.
This does not change the critical behavior, in particular it gives the same critical exponents.}
Then, with $p$ as the order parameter and $Q$ as the control variable,
 $\beta$ and $\delta$ are defined by
\beq |p| \sim |\eta|^\beta,\eeq
in analogy with the usual definition of $\beta$, but with $t$ replaced by $\eta$ and $p$ as the order parameter (so $p$ and $v$ are interchanged).
Similarly
\beq |v| \sim |p|^\delta \qquad \hbox{for} \qquad \eta =0. \eeq
Lastly the relevant response function is 
\beq -V\left.\frac{\partial P}{\partial V}\right|_{T,Q} = (\kappa_{T,Q})^{-1}.\eeq
$\gamma$ then follows from
\beq 
(\kappa_{T,Q})^{-1} \sim |\eta|^{-\gamma}.\eeq

We leave the details to an appendix and here quote the result that $\beta$, $\gamma$ and $\delta$ are indeed mean field.  We conclude that, with the proper identification of the order parameter as being the thermodynamic variable that jumps across the line of first order phase transformations, the exponents are always mean field in the large $N$ limit, both in the $\Phi$-$Q$ plane and the $P$-$V$ plane.

\section{Summary}\label{summary}

The critical point of ${\cal N}=4$ SUSY $SU(N)$ Yang-Mills in the large $N$ 
limit, first found in \cite{CEJM1,CEJM2},
is also visible in the $P-V$ diagram of the quark-gluon plasma. The volume here is provided by the cosmological constant in the $5$-D bulk, $\Lambda = -\frac{6}{L^2}$, which is related to the volume of $3$-space, $S^3$, in the boundary Yang-Mills theory by $V=2\pi^2 L^3$.  The pressure is then defined through the thermodynamic relation 
$P=-\frac{\partial U}{\partial V}$, where $U$ is the internal energy of the thermodynamic system --- identified with the mass of the black hole in the bulk.  This is in contrast to the thermodynamics of the black hole itself, where the roles of pressure and volume are reversed so that, on the gravity side, 
the cosmological constant provides the pressure and the thermodynamically conjugate variable is a volume and the black hole mass is the enthalpy of the system \cite{KRT}.

There is a constraint on the thermodynamic variables, due to the fact that the boundary field theory is conformal, and only the combination $VT^3$ is relevant to the phase structure, not $V$ and $T$ separately.  As a consequence it seems better to use the charge rather than the temperature as the control parameter in discussing the thermodynamics.
Then the pressure jumps across the phase transition in the two phase regime. In the approximation in which the fluid is treated as a gas of non-interacting 
particles -- quarks and gluons in the plasma and non-interacting hadrons in the confined phase --- this jump may be viewed as simply due
to a change in the number of degrees of freedom: each degree of freedom contributes equally to the pressure any change across the phase transition in the number of effective degrees of freedom contributes to a change in pressure. 

With pressure as the order parameter in the $P$-$V$ plane the exponents are mean field and the phase transition is in the same universality class as the van der Waals gas. Mean field exponents also characterize the phase transition in the $\Phi$-$Q$ plane.  
 
An unsatisfactory feature of the analysis is the instability associated
with the negative value of $C_P$ at the critical point, implying that the system is unstable there when the pressure is fixed.
Although $\kappa_T$ is positive in the region enclosed by the black curve to the right of the critical point in figure \ref{PV-figure} both $C_P$ and $C_V$ are negative there. The only region of the $P-V$ plane that has all three of $C_P$, $C_V$ and $\kappa_T$ positive is the region to the left of the left-hand black curve in the figure, the single phase region. $C_P<0$ to the right of this curve.
 
Instability in black hole thermodynamics is not new
and indeed lies behind the phenomenon of black hole evaporation --- an asymptotically flat Schwarzschild black hole also has negative $C_P$.
Whether or not negative $C_P$ is a problem is a question
of time-scales, for Hawking radiation the time-scale for black hole evaporation
can be very large, so large that the system is essentially quasi-static equilibrium and thermodynamic principles can still be applied, at least at early times before the evaporation process has gone too far. It is not clear whether
the same thing can be said for the quark-gluon plasma, or what the physical 
interpretation of the instability should be in this case.

Although the analysis here has focused on the case of 
4-dimensional ${\cal N}=4$ SUSY $SU(N)$ Yang-Mills, 
associated with 
string theory in ${\cal D}=10$ compactified on $AdS_5\times S^5$,
the thermodynamics of the 3-dimensional and 6-dimensional
CFTs, obtained from M-theory in
${\cal D}=11$ compactified on $AdS_4\times S^7/k$ and 
on $AdS_7\times S^4$ respectively,
is essentially similar --- the critical exponents are mean field in all three cases. 

It would be of great interest to evaluate $1/N$ corrections to the exponents but the method employed here relies on a representation of the thermodynamic potentials arising from an exact solution of the Einstein equations in the bulk. The paucity of known exact solutions relevant to the problem is an obstacle to realizing $1/N$ corrections with this method.  However it may be possible to make some progress in studying $1/N$ corrections by adding a Gauss-Bonnet term to the $5$-D Einstein action \cite{1overN}.

\bigskip

{\bf Acknowledgment:} This article is based upon work from COST Action MP1405 QSPACE, supported by COST (European Cooperation in Science and Technology).

\appendix

\section{Free energy calculations \label{AA}}

To study the theory at fixed $T$ in more detail consider the free energy
\beq F(T,V,Q)= U(S,V,Q) - TS,
\eeq
which is of the form
\beq F=\frac{\hbar N^2}{V^{\frac{1}{3}}} {\cal F}(\tau,y)
\eeq
where 
\beq
\tau:=\frac{ V^{\frac{1}{3}} T}{\hbar}
\eeq 
and ${\cal F}$ is a dimensionless function
of dimensionless variables. The free energy is most easily expressed in parametric form using
(\ref{Volume}) and (\ref{Temperature}),
\beq \label{fdef}
{\cal F} = \frac{(2\pi^2)^{\frac 1 3}}{4}\left(-x^4 + x^2 +\frac{5y^2}{x^2}\right),
\eeq
and
\beq \label{taudef}
\tau =  \frac{1}{(4\pi)^{\frac 1 3}}\left(2 x + \frac{1}{x} -\frac{y^2}{x^5}\right).
\eeq
In terms of the functions ${\cal F}(x,y)$ and $\tau(x,y)$ in equations
(\ref{fdef}) and (\ref{taudef}), 
\beq S=-\left.\frac{\partial F}{\partial T}\right|_{V,Q}
=-\frac{N^2}{V^{\frac 1 3}} \left.\frac{\partial \tau}{\partial T}\right|_{V,Q}
\frac {\partial f}{\partial \tau} = -N^2 \dot {\cal F},
\eeq
where $\dot {\cal F} = \frac{\partial {\cal F}}{\partial \tau}$, and
\beq 
\left.\frac{\partial S}{\partial T}\right|_{V,Q}
=-\left.\frac{\partial \tau}{\partial T}\right|_V N^2 \ddot {\cal F} = -V^{\frac 1 3} N^2 \ddot {\cal F}.
\eeq
%whence 
%\beq
%C_{V,Q}= T\left. \frac{\partial S}{\partial T}\right|_{V,Q}=N^2 \tau \ddot f.
%\eeq
Note also that
\beq
\left.\frac{\partial S}{\partial V}\right|_{T,Q}
=-N^2 \left.\frac{\partial \tau}{\partial V}\right|_T \ddot {\cal F} = -\frac {N^2} {3 V}\tau\ddot {\cal F}.
\eeq
From this follows
\beq C_{V,Q}=T\left.\frac{\partial S}{\partial T}\right|_{V,Q}= 
3V\left.\frac{\partial S}{\partial V}\right|_{T,Q} = -N^2 \tau\ddot {\cal F}.
\eeq
Now, since $P=\frac{U}{3V}$, we have
\bea \kappa_{T,Q}^{-1}=
-V\left.\frac{\partial P}{\partial V}\right|_{T,Q}&=& 
\frac{U}{3V} -\frac{1}{3}\left.\frac{\partial U}{\partial V}\right|_{T,Q}\nonumber \\
&=& P -\frac{1}{3} \left( \left.\frac{\partial U}{\partial V}\right|_{S,Q} 
+\left.\frac{\partial U}{\partial S}\right|_{V,Q}\left.\frac{\partial S}{\partial V}\right|_{T,Q}\right)\nonumber\\
&=& \frac{4P}{3}  - \frac{T C_{V,Q}}{9V}.
\eea
Thus the isothermal compressibility, $\kappa_{T,Q}$ vanishes at the critical point.  

\section{Critical behavior of pressure, volume and isothermal compressibility}
To derive the critical exponents for pressure and volume, $\beta$, $\gamma$ and $\delta$, first express the temperature and the pressure in reduced variables,
$t=\frac{T}{T_*}-1$ and $p=\frac{P}{P_*}-1$ at constant volume, 
equation (\ref{Temperature}) reads, with $x=x_*(1+\varepsilon)$,
\beq \label{ReducedTempRelation}
\eta(2+\eta)  + 24 t (1+\varepsilon)^5   = 20\varepsilon^3 + 45\varepsilon^4 + 36\varepsilon^5 + 10\varepsilon^6,
\eeq 
while eliminating $y=y_*(1+\eta)$ in favor of $T$ in (\ref{Pressure}) gives
\beq \label{ReducedPressure}
7 p + 8 t (1+\varepsilon)^3 = 16 \varepsilon + 16 \varepsilon^2 +12\varepsilon^3
+ 5 \varepsilon^4.
\eeq
With the definition of reduced volume (\ref{VT3}),
\beq v = (1+t)^3 -1, \eeq
equations (\ref{ReducedTempRelation}) and (\ref{ReducedPressure}) can be re-arranged
to give explicit expressions for $v(\varepsilon,\eta)$ and $p(\varepsilon,\eta)$
which, when expanded around the critical point, yield
\bea 
v &=& -\frac 1 4 \eta + \frac 5 4 \eta \varepsilon + \frac 5 2 \varepsilon^3
+\hbox{o}(\eta^2,\eta\varepsilon^2,\varepsilon^4)\label{eosv}\\
p& =& \frac 2 {21}\eta  - \frac 4 {21} \varepsilon \eta + \frac {16} {7}\varepsilon  + \frac{16}{7} \varepsilon^2 +\frac{16}{21}\varepsilon^3
+\hbox{o}(\eta^2,\eta^2\varepsilon,\varepsilon^4).\label{eosp}
\eea

\begin{figure}[ht]
\centerline{\includegraphics[scale=1]{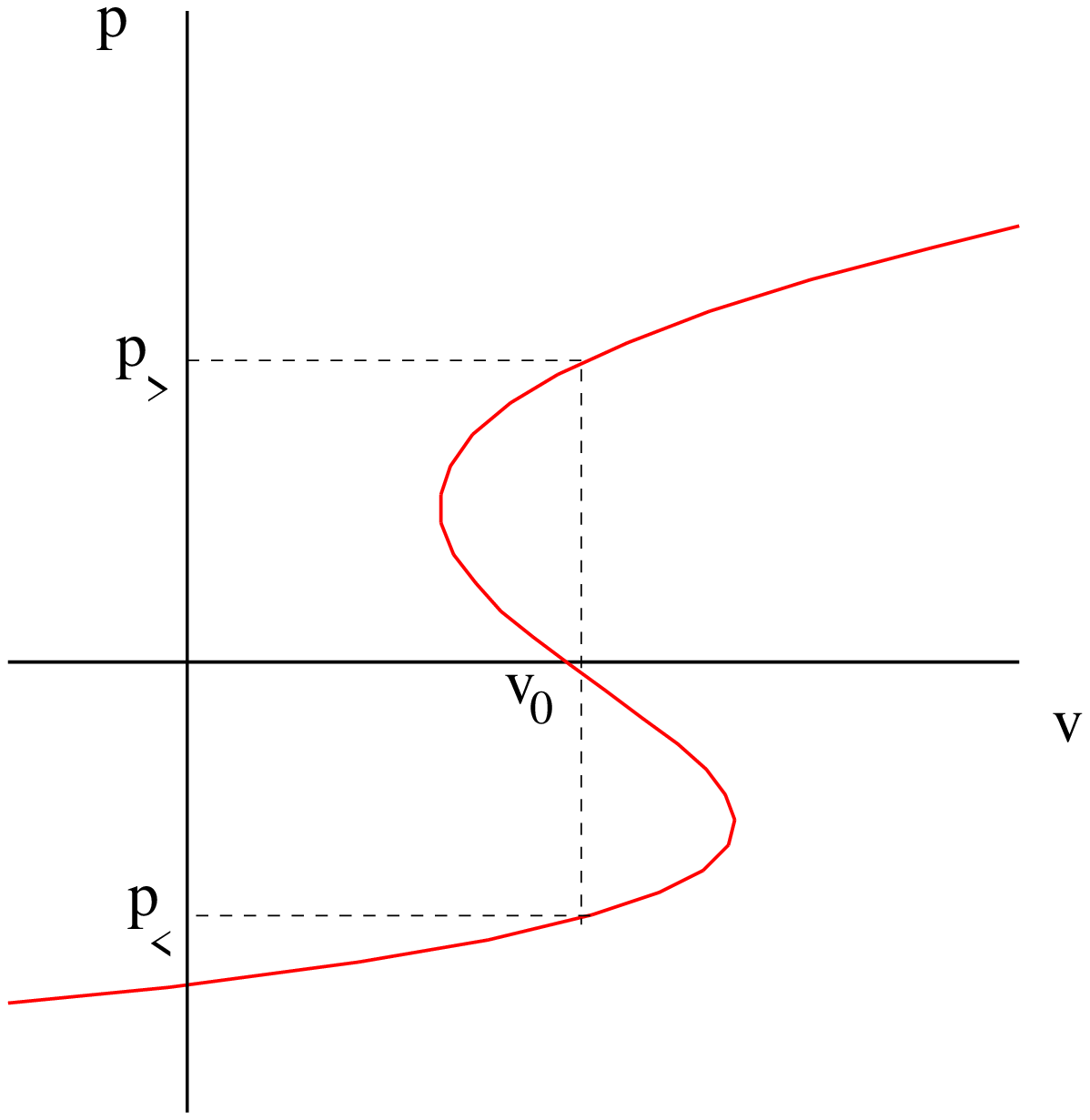}}
\caption{Maxwell construction in the $P$-$V$ plane along a curve of constant charge with $P$ as the order parameter ($P$ and $V$ are rendered dimensionless using appropriate powers of $T$ as in figure \ref{PV-figure}).}\label{Maxwell-construction}
\end{figure}

With $p$ as the order parameter and $\eta$ as the control parameter
the Maxwell construction demands that
the integral of $v(p,\eta)$ along a curve of constant $\eta$, between two values $p_<$ and $p_>$ on either side of the critical point with $v(p_<) = v(p_>)=v_0$, should satisfy
\beq \int_{p_<}^{p_>} v(p) d p - (p_> - p_<)v_0 =0\eeq
(see figure \ref{Maxwell-construction}).
 This requires
\beq \int p d v  = \int p(\varepsilon)\left.\frac{\partial v}{\partial \varepsilon}\right|_\eta  d\varepsilon \eeq
to have the same value at the two endpoints of integration, $p_< = p(\varepsilon_<)$ and $p_> = p(\varepsilon_>)$.
Explicitly, expanding (\ref{eosv}) and (\ref{eosp}),
\beq
\int p(\varepsilon)\frac{\partial v}{\partial \varepsilon} d\varepsilon 
= \frac{10}{7} \eta\,\varepsilon^2+ \frac{5}{42}  \eta^2  \varepsilon + \frac{30}{7}\varepsilon^4 + \hbox{o}(\eta\,\varepsilon^3,\eta^3\varepsilon,\varepsilon^5).\eeq If $\varepsilon_< = -\varepsilon_>$ then even powers of $\varepsilon$ cancel in the definite integral 
\beq
\int_{\varepsilon_<}^{\varepsilon_>}  
p(\varepsilon)\frac{\partial v}{\partial \varepsilon} d\varepsilon = 
\frac{5}{21} \eta^2\varepsilon_> + 
\hbox{o}{(\eta\,\varepsilon_>^3,\eta^3\varepsilon_>,\varepsilon_>^5)}.
\eeq
Equation (\ref{eosv}) implies that $v(\epsilon_<) = v(\varepsilon_>) + $ o$\,(\varepsilon^4)$ if $\eta =-2\varepsilon_<^2=-2\varepsilon_>^2$, in which case the integral is o$\,(\varepsilon_>^5)$.
Using this in (\ref{eosp}) finally produces
\beq p \sim  |\eta|^{\frac 1 2} + \hbox{o}|\eta|\qquad \Rightarrow \qquad \beta = \frac 1 2.
\eeq

The exponent $\gamma$ is obtained from the isothermal compressibility in equation (\ref{kappaT}) (equation (\ref{ReducedPressure}) should not be used for differentiation here as $p$ is defined at fixed volume, not at fixed temperature)
\beq
\kappa_{T,Q}=-{\frac {12{\pi }^{2}{L}^{4} \left( 2\,{x}^{6}-{x}^{4}+5\,{y}^{2}
 \right) {x}^{2}}{{N}^{2} \left( 2\,{x}^{10}-18\,{x}^{6}{y}^{2}+3\,{x}
^{8}-10\,{x}^{4}{y}^{2}-9\,{y}^{4} \right) }}.
\eeq
Expanding this around the critical
point one finds that $\kappa_{T,Q}$ vanishes like
\beq
\kappa_{T,Q} \propto -\eta - 6\varepsilon^2 + \hbox{o}(\eta^2,\varepsilon^3) 
= 12 t(1+5\varepsilon) - 6\varepsilon^2 + \hbox{o}(t^2,t\varepsilon^2,\varepsilon^3) 
\eeq
where (\ref{ReducedTempRelation}) has been used to eliminate $\eta$
in the second equation.
Now we can set $p=0$ in (\ref{ReducedPressure}) and equation 
(\ref{ReducedTempRelation})
then implies 
$\varepsilon \approx \frac{1} {2} t$
and $\eta \sim -24\varepsilon$, so
\beq
\kappa_{T,Q} \sim  -\eta + \hbox{o}(\eta^2) \sim 12 t + \hbox{o}(t^2). 
\eeq
Hence $(\kappa_{T,Q}){}^{-1} \sim 1/|\eta| \sim 1/t$ at the critical pressure and, with pressure as the order parameter, $\gamma=1$. In the 2-phase regime ($\eta <0$) the isothermal compressibility is positive, but it is negative for $\eta=0$.

Lastly setting $\eta=0$ in the parametric equation of state (\ref{eosv}) and (\ref{eosp}) implies that
$p\approx \frac{16}{7} \varepsilon$ and  $v\approx \frac 5 2 \varepsilon^3$
so
\beq
p \sim v^{\frac 1 3} + \hbox{o}(v^{\frac 2 3})\qquad \Rightarrow \qquad v \sim p^3
\eeq
and hence $\delta =3$.


\begin{thebibliography}{15}

\bibitem{Bekenstein} J.D.~Bekenstein, {\it Lett.~Nuovo.~Cimento} 
{\bf 4} (1972) 737;\hfill\break
J.D.~Bekenstein, {\it Phys. Rev. D} {\bf 7} (1973) 2333.
% Area and entropy

\bibitem{Hawking} S.W.~Hawking, {\it Nature} {\bf 248} (1974) 30;\hfill\break
S.W.~Hawking, {\it Comm.~Math.~Phys.} {\bf 43} (1975) 199;\hfill\break 
S.W.~Hawking, {\it Phys. Rev. D} {\bf 13} (1976) 191.
% Black hole temperature and S=(1/4)A.

\bibitem{ABJM}  O.~Aharony, O.~Bergmann, D.L.~Jafferis and J.~Maldacena, {\it JHEP} {\bf 10} (2008) 091, [arXiv:0806.1218].

\bibitem{Henneaux} M.~Henneaux and C.~Teitelboim, {\it Phys.~Lett.} {\bf 143B}, (1984) 415; {\it ibid.} {\bf 222B}, (1989) 195.
%

\bibitem{KRT} D.~Kastor, S.~Ray and J.~Traschen,
%{\it Enthalpy and the Mechanics of AdS Black Holes},
{\it Class. Quantum Grav.} {\bf 26}, (2009) 195011, [arXiv:0904.2765].

\bibitem{Galaxies}  N.~Altamirano, D.~Kubiz\u{n}\'ak, R.B.~Mann and Z.~Sherkatghanad, \newline
%{\sl Thermodynamics of rotating black holes and black rings: phase transitions and thermodynamic volume,}
{\it Galaxies} {\bf 2} (2014) 89, [arXiv:1401.2586].

\bibitem{LambdaReview} B.P.~Dolan,
%{\sl  Black holes and Boyle's law -- the thermodynamics of the cosmological constant}
{\it Mod. Phys. Lett.} {\bf A30} (2015) 1540002, [arXiv:1408.4023].

\bibitem{CJ} C.V.~Johnson, {\it Class. Quant. Grav.} {\bf 31} (2014) 205002,
[ arXiv:1404.5982].

\bibitem{BCB} B.P.~Dolan, {\it JHEP} {\bf 10} (2014) 179, [arXiv:1406.7267]. 

\bibitem{KR} A.~Karch and B.~Robinson, %{\sl Holographic Black Hole Chemistry}, 
{\it JHEP} {\bf 12} (2015) 073, [arXiv:1510.02472].

\bibitem{HP} S.W.~Hawking and D.N.~Page,
% Thermodynamics of black holes in anti-de Sitter space
{\it Comm. Math. Phys.} {\bf 87}, (1983) 577.

\bibitem{Witten} E.~Witten, 
{\it Adv. Theor. Math. Phys.} {\bf 2} (1998) 253, [hep-th/9802150];
{\it ibid.} {\bf 2} (1998) 505, [hep-th/9803131].

\bibitem{CEJM1} A.~Chamblin, R.~Emparan, C.V.~Johnson and R.C.~Myers,\hfill\break
%{\sl Charged AdS Black Holes and Catastrophic Holography}
{\it Phys. Rev. D} {\bf 60} (1999) 064018, [hep-th/9902170].

\bibitem{NTW} C.~Niu, Y.~Tian and X.~Wu, {\it Phys. Rev. D} {\bf 85} (2012) 024017, 
[arXiv: 1104.3066].

\bibitem{Maldacena} J.~Maldacena, {\it Adv. Theor. Math. Phys.} {\bf 2} (1998) 231, [hep-th/9711200].

\bibitem{KM} D.~Kubiz\u{n}\'ak and R.B.~Mann, 
% P-V criticality of charged AdS black holes
{\it JHEP} {\bf 1207} (2012) 033, [arXiv:1205.0559]. 

\bibitem{CEJM2} A.~Chamblin, R.~Emparan, C.V.~Johnson and R.C.~Myers,\hfill\break
%{\sl Holography, Thermodynamics and Fluctuations of Charged AdS Black Holes}
{\it Phys. Rev.D} {\bf 60} (1999) 104026, [hep-th/9904197].

\bibitem{CCK} M.M.~Caldarelli, G.~Cognola and D.~Klemm, 
{\it Class. Quantum Grav.} {\bf 17} (2000) 399, [arXiv:hep-th/9908022].

\bibitem{Stability} B.P.~Dolan, 
%{\it Thermodynamic stability of asymptotically anti-de Sitter rotating black holes in higher dimensions },
{\it Class. Quantum Grav.} {\bf 31} (2014) 165011, [arXiv:1403.1507].

%\bibitem{Veselin} T.~Albash, V.~Filev, C.V.~Johnson and A.~Kundu, 
%{\it JHEP} {\bf 12} (2008) 033, [hep-th/0605175].

\bibitem{1overN} S.~Nojiri and S.D.~Odintsov, {\it JHEP} {\bf 0007} (2000) 049, [hep-th/0006232].

\end{thebibliography}
\end{document}